\documentclass[aps,pra,twocolumn,amsfonts,amssymb,amsmath,showpacs,
floatfix,nofootinbib,groupedaddress,superscriptaddress,citesort]{revtex4}
\usepackage{mathrsfs}
\usepackage{amsfonts}
\usepackage{amstext}
\usepackage{amsmath}
\usepackage{amssymb}
\usepackage{bm}% bold math
\usepackage{bbm}
\usepackage[dvips]{graphicx}
\def\qed{\leavevmode\unskip\penalty9999 \hbox{}\nobreak\hfill
     \quad\hbox{\leavevmode  \hbox to.77778em{%
              \hfil\vrule   \vbox to.675em%
               {\hrule width.6em\vfil\hrule}\vrule\hfil}}
     \par\vskip3pt}

\begin{document}
\title
{State convertibility under genuinely incoherent operations}

\author{Shuanping Du}
\email{dushuanping@xmu.edu.cn} \affiliation{School of Mathematical
Sciences, Xiamen University, Xiamen, Fujian, 361000, China}

\author{Zhaofang Bai}\thanks{Corresponding author}
\email{baizhaofang@xmu.edu.cn} \affiliation{School of Mathematical
Sciences, Xiamen University, Xiamen, Fujian, 361000, China}

%\thanks{{\it PACS numbers: 03.67.HK, 02.10.Yn, 03.65.Aa}}
%\thanks{{\it Mathematics Subject Classification: 47B49, 46L07, 47L07, 46N50}}
%\thanks{{\it Key words and phrases.} strictly incoherent operations, frozen coherence, the $l_1$-norm of coherence
%}

%\thanks{This paper is in final form and no version of it will be
%submitted for publication elsewhere.}

\begin{abstract}

State convertibility is fundamental in the study of resource theory of quantum coherence. It is aimed
at identifying when it is possible to convert a given coherent state to another using only incoherent operations. In this paper, we give a complete characterization of  state convertibility under genuinely incoherent operations. It is found that convexity of the robustness of coherence plays a central role. Based on this, the majorization condition of determining convertibility from pure states to mixed states under strictly incoherent operations is provided. Moreover,  maximally coherent states in the set of all states with fixed diagonal elements are determined.
It is somewhat surprising that convexity of the robustness of coherence can also decide
conversion  between off-diagonal parts of coherent states. This might be a big step to answer completely the question of state convertibility  for mixed states under incoherent operations.
\end{abstract}

\pacs{03.65.Ud, 03.67.-a, 03.65.Ta.}

\maketitle

\section{Introduction }

The coherent superposition of states
is one of the characteristic features that results in nonclassical phenomena \cite{Legg,Mandel}.
Quantum coherence constitutes a powerful physical resource for implementin
various tasks such as quantum algorithms \cite{Hillery,Matera,Pan1,Wang,Annefeld,Karimi,Ye,Berberich,Moreno}, quantum metrology \cite{Giovannetti,Giovannetti2,Demkowicz, Pires,Cheng,Zhang,Castellini,Ares,Lecamwasam}, quantum channel discrimination \cite{Girolami,Farace,Streltsov,Takagi,Wilde,Rossi,Chen2}, witnessing quantum correlations \cite{Ma,Hu,Hu2,Mondal,Girolami2,Ding,Lee}, quantum phase transitions and transport phenomena \cite{Karpat,Cakmak,Malvezzi,Chen3,Li,Shi}. The resource theory of quantum coherence has been flourishing in recently years, it not only establishes a rigorous framework to
quantify coherence but also provides a platform to understand quantum coherence from a different perspective \cite{Baumgratz,Streltsovade}.

Any quantum resource theory is described by two fundamental ingredients, namely, the free states and the free
operations \cite{Chitambar}. For the resource theory of coherence, the free
states are quantum states which are diagonal in a prefixed
reference basis. The free operations are not uniquely specified.
Motivated by different physical considerations, several free
operations are presented , such as incoherent operations (IOs) \cite{Baumgratz}, maximally incoherent
operations (MIOs) \cite{Aberg},
strictly incoherent operations (SIOs) \cite{Winter,Yadin}, dephasing-covariant
incoherent operations (DIOs) \cite{Chitambar2,Chitambar3,Marvian}, and genuinely incoherent operations (GIOs) \cite{Vice}.
%In fact, there is a hierarchical relationship between IOs, MIOs, SIOs, DIOs and GIOs, $\text{GIOs}\subseteq \text{SIOs}\subseteq \text{IOs}\subseteq \text{MIOs}$, $\text{GIOs}\subseteq \text{SIOs}\subseteq \text{DIOs}\subseteq \text{MIOs}$ \cite{Streltsovade}.

Two fundamental problems in coherence resource
theory are state convertibility and resource quantification \cite{Streltsovade,Chitambar}.
The state convertibility problem is asking whether for two
coherent states there exists a free operation converting one quantum
state into the other. The goal of resource quantification is to
quantify the amount of the coherence in a quantum state.  Recalling that
coherent states cannot be created from incoherent states via free
operations, it is intuitive to assume that $$C(\rho)\geq C(\Phi(\rho))\ \eqno{(1)}$$
for any quantum state $\rho$ and any free operation $\Phi$.
Quantifiers having this property are also called coherence monotones.

Both problems mentioned above--- state convertibility
and resource quantification are in fact closely connected.
A state $\rho$ can be converted into $\sigma$ via free operations if and
only if $$C(\rho)\geq C(\sigma)\ \eqno{(2)}$$ holds true for all coherence monotones \cite{Takagix}.
On the other hand, the fact that Eq. (2) holds for some coherence monotone $C$ does not guarantee that the transformation $\rho \rightarrow\sigma$ is
possible via free operations. The aim of state convertibility is to find a
complete set of coherence monotones $\{C_i\}$ which can completely classify states transformation, i.e.,
$$\rho\rightarrow \sigma\Leftrightarrow C_i(\rho)\geq C_i(\sigma)\ \eqno{(3)}$$ for all $i$.

%The main question of quantum coherence
% is to order the set of states and provide means
%to quantify their coherence as a resource. This raises an interesting question named state convertibility \cite{Baumgratz}.
%The state convertibility  is asking whether for two
%coherent states there exists a free operation converting one
%state into the other. If a state $\rho$ can be
%transformed into $\sigma$ by some free operation, then $\rho$ cannot
%be less resourceful than $\sigma$ since any task achievable by $\sigma$ is
%also achievable by $\rho$ as the corresponding transformation
%can be freely implemented.
The study of state convertibility is moving ahead since the question is proposed \cite{Baumgratz}, it is completely answered in pure states or one-qubit case under IOs, SIOs or MIOs \cite{Winter, Chitambar2,Du1,Streltsov3,Zhao2,Torun1,Torun2,Lami2,Du2,Liu2,Wu,Fang,Torun3,Gaot,Torun4}. The convertibility between  mixed
 states seems to have remained unexplored territory.  The difficulty lies in the complexity of pure state decomposition
which results in infinite number measure conditions for characterizing convertibility of mixed states \cite{Du2}.
We investigate convertibility for coherent states under
GIOs. In fact, GIOs are at the core of the resource theory of quantum
coherence from both physical realization and dissecting the structure of SIOs and IOs \cite{Sun}.
Note that there is  a hierarchical relationship between IOs, SIOs, MIOs, and GIOs \cite{Streltsovade}, $$\text{GIOs}\subseteq \text{SIOs}\subseteq \text{IOs}\subseteq \text{MIOs}.\ \eqno{(4)}$$
 For any $\mathcal O\in \{\text {IOs, SIOs,\ MIOs}\}$, we define $\rho \xrightarrow{{\mathcal O}}\sigma$ if there exists $\Phi\in\mathcal O$ such that $\Phi(\rho)=\sigma$. It is evident if $\rho \xrightarrow{\text{GIO}}\sigma$, then $\rho \xrightarrow{{\mathcal O}}\sigma$.
%This implies state convertibility under GIOs constitutes the core.

 %It has been shown that. In fact, there is a. Therefore.

 %The convertibility between coherent mixed
 %states seems to have remained unexplored territory.  The difficulty lies in the complexity of pure state decomposition
%which results in infinite number measure conditions for characterizing convertibility of mixed states \cite{Du2}. On the other hand, noise renders all states mixed  in practice and pure
%states are an idealization, hence the study of convertibility for mixed states is pressing.

 The complete set of coherence monotones for characterizing state convertibility under GIOs are found. In fact, convexity of the robustness of coherence is a good candidate. Moreover, it is also key to convert off-diagonal part of coherent states under more general free operations. Our results  induce a useful tool for deciding maximally coherent states in the set of all states with fixed diagonal elements. This produces so-called  majorization condition of determining convertibility from
pure states to mixed states under SIOs.

The paper is organized as follows. In section II, we briefly present the resource theory of quantum coherence. In section III, we will give our main results. Section IV is a summary of our findings. The appendix is the proof of our results.

\section{Definition and basic properties}

Throughout the paper, we consider the $d$ dimensional Hilbert space ${\mathcal H}$ and adopt the computational basis $\{|i\rangle\}_{i=1}^d$ as the incoherent basis \cite{Baumgratz}. Thus all diagonal density operators in this basis constitute the set of all incoherent states denoted as  ${\mathcal I}$. IOs are specified by a set of Kraus operators $\{K_j\}$ such that $K_j\rho K_j^\dag/Tr(K_j\rho K_j^\dag)\in {\mathcal
	I}$ for all $\rho\in {\mathcal I}$, $\Phi(\rho)=\sum_{j}K_j \rho K_j^\dag.$
Such operation elements $\{K_j\}$ are called incoherent. An
incoherent operation is strictly incoherent  if both $K_j$ and $K_j^\dag$
are incoherent.
The MIOs  are known as incoherent states preserving operations.
GIOs are operations which fix all
incoherent states, i.e., $$\Phi(\rho)=(\rho)\ \eqno{(5)}$$ for any incoherent state $\rho\in{\mathcal I}$.  Since GIOs do not allow for
transformations between different incoherent states, notably,
for example, between the energy eigenstates (when coherence
is measured with respect to the eigenbasis of the Hamiltonian
of the system), they capture the framework of coherence in the
presence of additional constraints, such as energy conservation. For other important type of incoherent operations,
we refer the reader to the review article \cite{Streltsovade}.
%t has been shown that GIOs in fact constitute the core of other types of incoherent operations \cite{Sun}. In fact, there is a hierarchical relationship between IOs, SIOs, MIOs, and GIOs, $\text{GIOs}\subseteq \text{SIOs}\subseteq \text{IOs}\subseteq \text{MIOs}$ \cite{Streltsovade}. Thus if $\rho \xrightarrow{\text{GIO}}\sigma$, then $\rho \xrightarrow{{\mathcal O}}\sigma$
%for any $\mathcal O\in \{\text {IOs, SIOs,\ MIOs}\}$. Therefore state convertibility under GIOs is more fundamental.

In order to characterize conversion of coherent states under GIOs,  we need a key measure originated from the task of maximizing the mean value of an observable \cite{Cktan}. Let $|\psi^+\rangle=\frac{1}{\sqrt d}\sum_{i=1}^d|i\rangle$, it is well-known that  $|\psi^+\rangle\langle \psi^+|$ is a maximally coherent state under IOs, i.e., a state from which all other states can be created via
IOs \cite{Baumgratz}. It is easy to see that $U|\psi^+\rangle\langle\psi^+|U^\dag$ is maximally coherent under IOs for any diagonal unitary matrix $U$. Let $\Omega$ be the set of convex hull of $U|\psi^+\rangle\langle\psi^+|U^\dag$.
For every $M\in \Omega$, define $$C_{ M}^\text{GIOs}(\rho)=\max_{\Phi\in\text{ GIOs}} {\rm tr}(\Phi(\rho)M)-\frac{1}{d}.\ \eqno{(6)}$$
In the following, we list some elementary properties of $C_{ M}^\text{ GIOs}(\cdot)$ and discuss its relationship with other coherence measures (see appendix for the proof).

(i) $C_{ M}^\text{GIOs}(\rho)\geq 0$ for every quantum state $\rho$ and $C_{ M}^\text{GIOs}(\rho)=0$ if $\rho\in{\mathcal I}$;

(ii) Monotonicity under all GIOs $\Phi$: $$C_{ M}^\text{GIOs}(\Phi(\rho))\leq C_{ M}^\text{GIOs}(\rho);\ \eqno{(7)}$$

(iii) Monotonicity for average coherence: $$\sum_jp_jC_{ M}^\text{GIOs}(\rho_j)\leq C_{ M}^\text{GIOs}(\rho)\ \eqno{(8)}$$ for all $\{K_j\}$ specifying every GIO,
where $\rho_j=\frac{K_j\rho K_j^\dag}{p_j}$ and $p_j={\rm Tr}(K_j\rho K_j^\dag)$;

(iv) Non-increasing under mixing of quantum states:
$$C_{ M}^\text{GIOs}(\sum_jp_j\rho_j)\leq \sum_jp_jC_{ M}^\text{GIOs}(\rho_j)\ \eqno{(9)}$$ for any set of states
$\{\rho_j\}$ and any $p_j\geq 0$ with $\sum_jp_j=1$;

(v) $C_{ M}^\text{GIOs}(\rho)$ is related to the $l_1-$norm of coherence by the inequality $$\begin{array}{ll}
\frac{C_{l_1}(\rho)}{d-1}\min_{1\leq i\neq j\leq d}\{|M_{ij}|\}&\leq C_{ M}^\text{GIOs}(\rho)\\
\leq C_{l_1}(\rho)\max_{1\leq i\neq j\leq d}\{|M_{ij}|\},& \end{array} \ \eqno{(10)}$$ here $M=(M_{ij})$ and $C_{l_1}(\rho)=\sum _{i\neq j}|\rho_{ij}|$ is the $l_1-$norm of coherence;

(vi) $C_{ M}^\text{GIOs}(\rho)$ is also related to the robustness of coherence by the inequality $$0\leq C_{ M}^\text{ GIOs}(\rho)\leq \frac{C_{\text ROC}(\rho)}{d},\ \eqno{(11)}$$
here $$C_{\text ROC}=\min_{\tau\in {\mathcal S}}\ \ \{s:\frac{\rho+s\tau}{1+s}\in{\mathcal I}\}=\min_{\delta\in{\mathcal I}}\{s:\rho\leq (1+s)\delta\}$$ is the robustness of coherence \cite{Napoli}.

Specially, if $M=|\psi^+\rangle\langle \psi^+|$, then
$$C_{ |\psi^+\rangle\langle \psi^+|}^\text{GIOs}(\rho)=\frac{C_{\text ROC}(\rho)}{d}\ \eqno{(12)}$$ \cite{Sun2}. %Note that $C_{ |\psi^+\rangle\langle \psi^+|}^\text{GIOs}(\rho)=C_{ U|\psi^+\rangle\langle \psi^+|U^\dag}^\text{GIOs}(\rho)$ for any diagonal unitary matrix $U$.
For general $M\in\Omega$, there exist a probability distribution $\{p_i\}$ and diagonal unitary matrices $\{U_i\}$ such that $M=\sum_i p_iU_i|\psi^+\rangle\langle\psi^+| U_i^\dag$. That is, $M$ is a convexity of maximally coherent states.  In this sense, we say $C_M^\text{GIOs}$
is a convexity of the robustness of coherence.
 It is found that such measures plays a key role for studying state convertibility under GIOs.

\vspace{0.1in}

\section{Main results}

Now, we are in a position to give our main result.

\vspace{0.1in}

\textbf{Theorem 3.1.} {\it There exists some $\text{GIO}$ $\Phi$ such that $$\Phi(\rho)=\sigma\Leftrightarrow C_M^{\text GIOs}(\rho)\geq C_M^{\text GIOs}(\sigma) \ \eqno{(13)}$$ $\text{ for any } M\in {\Omega}, \ \rho_{ii}=\sigma_{ii} \ (i=1,2,\dots, d).$}

\vspace{0.1in}

Theorem 3.1 tells convertibility between pure states is impossible except for diagonal-unitary equivalent states. A parallel result in  multipartite entanglement is almost all $n$-qubit pure states with $n\geq 3$ can neither be reached nor be converted into any other LU-inequivalent state via deterministic LOCC \cite{Sauerwein}. On the other hand, deterministic convertibility between incoherent-unitary inequivalent pure states is possible under IOs, DIOs, SIOs, and MIOs \cite{Du1,Chitambar2}. Thus, compared with other free operations in the coherence resource theory, GIOs are more matching to LOCC in multipartite entanglement theory from the point of state convertibility  .

For one-parameter maximally mixed states \cite{Streltsov3,Sun2}$$\rho_p=p|\psi^+\rangle\langle\psi^+|+\frac{1-p}{d}I, \ \eqno{(14)}$$ Theorem 3.1 shows that  $$\rho_p \xrightarrow{\text{GIO}}\rho_q\Leftrightarrow q\leq  p.\ \eqno{(15)}$$

Based on Theorem 3.1, we can provide a nice majorization condition which determines the convertibility from pure states to mixed states under SIOs, IOs, and MIOs.

\vspace{0.1in}

\textbf{Theorem 3.2.} {\it For $|\psi\rangle=\sum_{i=1}^d\psi_i|i\rangle$, $\sigma=(\sigma_{ij})$, $$(|\psi_1|^2, \cdots, |\psi_d|^2 )^t\prec (\sigma_{11}, \cdots,\sigma_{dd})^t \Rightarrow|\psi\rangle\langle\psi| \xrightarrow{\text{SIO}}\rho\ \eqno{(16)}$$ here $\prec$ denotes the majorization relation between probability vectors.}

\vspace{0.1in}
By the hierarchical relationship $\text{SIOs}\subseteq \text{IOs}\subseteq \text{MIOs}$,  there exists some IO or MIO $\Phi$ with
$\Phi(|\psi\rangle\langle\psi|)=\sigma$ if $(|\psi_1|^2, \cdots, |\psi_d|^2 )^t\prec (\sigma_{11}, \cdots,\sigma_{dd})^t$.

For $|\psi^+\rangle=\sum_{i=1}^d \frac{1}{\sqrt d}|i\rangle,$ it is evident that $$(\frac{1}{d}, \cdots, \frac{1}{d} )^t\prec (\sigma_{11}, \cdots,\sigma_{dd})^t\ \eqno{(17)}$$ for any quantum state $\sigma.$ A direct consequence of Theorem 3.2 is that $|\psi^+\rangle\langle\psi^+|$ is maximally coherent under IOs which is an important conclusion of \cite{Baumgratz}.

It is well-known that convertibility between pure states is completely characterized by majorization relation \cite{Du1}. Theorem 3.2 can be regared as an extension when the output state is mixed. Although a structural
characterization of coherence conversion for the output mixed state is provided in terms of a finite number of measure conditions \cite{Du2},
such conditions are somewhat hard to verify because pure state decomposition is involved.
In comparison, Theorem 3.2 is more handy because we need only to check a majorization relation.

\vspace{0.1in}

The core for the proof of Theorem 3.2 is to find maximally coherent states (MCS) in the set of all states ${\mathcal S}$ with fixed diagonal elements, here a MCS means a state from which all other states of ${\mathcal S}$ can be created via GIOs.

We remark that the existence of MCS in a particular set of states ${\mathcal S}$ has independent meaning, because  one
may not be able to prepare all states of choice in many situations. Suppose we are bound to a particular set of states ${\mathcal S}$, can we find a notion of maximally coherent state in ${\mathcal S}$.  By Theorem 3.1, a natural choice of ${\mathcal S}$
is the set of all states with fixed diagonal elements, i.e., ${\mathcal S}=\{(\rho_{ij}): \rho_{ii}=p_i, i=1, 2,\cdots, d\}$, here $\{p_i\}$ is a fixed probability distribution.  In fact, there exists a MCS in ${\mathcal S}$. Our result reads as follows.

\vspace{0.1in}

\textbf{Theorem 3.3.} {\it  Let  $|\psi\rangle=\sum_{i=1}^{d}\sqrt p_i |i\rangle$, ${\mathcal S}=\{(\rho_{ij}): \rho_{ii}=p_i, i=1, 2,\cdots, d\}$. Then for any $\rho\in{\mathcal S}$, there exists a $\text {GIO}$ $\Phi$ such that $\Phi(|\psi\rangle\langle\psi|)=\rho$.}

\vspace{0.1in}

By the hierarchical relationship $$ \text{GIOs}\subseteq \text{SIOs}\subseteq \text{IOs}\subseteq \text{MIOs},$$
$$ \text{GIOs}\subseteq \text{SIOs}\subseteq \text{DIOs}\subseteq \text{MIOs},$$ we can obtain $|\psi\rangle=\sum_{i=1}^d \sqrt{p_{i}}|i\rangle$
is also maximally coherent in ${\mathcal S}$ under  SIOs, DIOs, IOs, and MIOs.

Theorem 3.1 and Theorem 3.3 shows that coherent mixed states can not be converted into pure states in general.
This is a parallel result of no-go theorem of purification for coherent mixed states of discrete-variable and Gaussian systems \cite{Fang,Du4}.
It shows a strong limit on the efficiency of perfect coherent purification under GIOs.

We also remark that parallel discussion of Theorem 3.3 in quantum entanglement is the existence of a maximally entangled state within a
given set of states with fixed spectrum. This is The
Problem 5 in the Open Quantum
Problems List maintained by the Institute for Quantum
Optics and Quantum Information (IQOQI) in Vienna \cite{Kruger,oqp}. It is newly shown that maximally entangled mixed states for a fixed spectrum do not always exist \cite{Julio}.

\vspace{0.1in}

%Theorem 3.1 and Theorem 3.2 demonstrate once again that there does not exist a maximally coherent
%state under GIOs, i.e., a state from which all other states can be created via GIOs \cite{Vice}. It is due to the fact that
%coherent states with different diagonal entries can neither be reached nor be converted into any other. Therefore $|\psi^+\rangle\langle\psi^+|$
%is not maximally coherent under GIOs although it is maximally coherent under SIOs, IOs or MIOs \cite{Streltsovade}.
%However, it is amazing that $|\psi^+\rangle\langle\psi^+|$ is in fact maximally coherent under GIOs in the set of coherent states with  diagonal elements $\frac{1}{d}$.

By Theorem 3.1,  if diagonal elements of $\rho$ and $\sigma$ are not completely equal in the same position, then both $\rho \nrightarrow\sigma$ and
$\sigma \nrightarrow\rho$ under GIOs hold true. However, exact conditions for realizing  conversion between off-diagonal parts of coherent states can also be found.

For any $\mathcal O\in \{\text {GIOs,\ DIOs\ MIOs}\}$,  we define $$C_M^{\mathcal O}(\rho)=\max_{\Phi\in\mathcal{O}} {\rm tr}(\Phi(\rho)M)-\frac{1}{d}\ \eqno{(19)}$$
for $M\in\Omega$. By the hierarchical
relationship between GIOs, DIOs and MIOs \cite{Streltsovade}, we know that each $C_M^{\mathcal O}(\cdot)$ is a coherence measure. Based on this, we actually have the following result.

\vspace{0.1in}

\textbf{Theorem 3.4.} {\it There exists some $\Phi\in \mathcal{O}$ such that $$\Phi(\rho)-\triangle (\Phi(\rho))=\sigma-\triangle(\sigma) \Leftrightarrow C_M^{\mathcal O}(\rho)\geq C_M^{\mathcal O}(\sigma), \ \eqno{(20)}$$ $\text{ for any } M\in {\Omega},$ here $\triangle$ is the dephasing operation defined by $\triangle(\rho)=\sum_{i=1}^d|i\rangle\langle i|\rho|i\rangle\langle i|$.}

\vspace{0.1in}

%One may ask why $C_M^{\mathcal O}(\rho)\geq C_M^{\text GIO}(\sigma)$ can not decide the conversion of diagonal part of coherent states, that is $C_M^{\mathcal O}(\rho)\geq C_M^{\text GIO}(\sigma)\nRightarrow\Phi(\rho)_{ii}=\sigma_{ii} (i=1,2,\cdots,d)$?

%By the hierarchical relationship $$ \text{GIOs}\subseteq \text{SIOs}\subseteq \text{IOs}\subseteq \text{MIOs},$$
%$$ \text{GIOs}\subseteq \text{SIOs}\subseteq \text{DIOs}\subseteq \text{MIOs},$$ we can obtain $|\psi\rangle=\sum_{i=1}^d \sqrt{p_{i}}|i\rangle$
%is also maximally coherent in ${\mathcal S}$ under  SIOs, DIOs, IOs, and MIOs.

%The related discussion in quantum entanglement is the existence of a maximally entangled state within a
%given set of states with the fixed spectrum. This is The
%Problem 5 in the Open Quantum
%Problems List maintained by the Institute for Quantum
%Optics and Quantum Information (IQOQI) in Vienna \cite{Kruger,oqp}. It is newly shown that maximally entangled mixed states for a fixed spectrum do not always exist \cite{Julio}.

Imaginarity as resource is a hot topic and recently receives much
attention (see \cite{Xu2} and the references therein).  For any coherence measure $C$, $$C(\rho)= C(\rho^*)\ \eqno{(21)}$$ is an axiomatic assumption proposed in \cite{Xu2} for studying coherence and imaginarity of quantum states, here $ \rho^*$ is the complex conjugate of $\rho$.
The intuition tells us (21) is right. Actually, the author has checked that all existing important coherence measures such as the $l_1$-norm of coherence, the relative entropy of coherence \cite{Baumgratz},  the Tsallis relative entropy of coherence \cite{Yuz}, the robustness of coherence,
the geometric coherence \cite{Streltsovx}, the coherence weight \cite{Bukai}, and coherence measures from the convex roof construction \cite{Du3}
 satisfying $C(\rho)=C(\rho^*).$ From the point of state convertibility, we need only to prove $$\rho\xrightarrow{\Phi_1}\rho^*,\ \rho^*\xrightarrow{\Phi_2}\rho ,\ \eqno{(22)}$$ $\Phi_1, \Phi_2\in\{\text{GIOs, DIOs, MIOs}\}$.
By Theorem 3.4, we need to check $C_M^{\mathcal O}(\rho)=C_M^{\mathcal O}(\rho^*)$. However, we find that $C_M^{\text{GIOs}}(\cdot)$ has a distinguished  property $$C_M^{\text{GIOs}}(\rho)\neq C_M^{\text{GIOs}}(\rho^*)\ \eqno{(23)}$$ for some $\rho$ and $M\in\Omega$ (see the appendix for an example).
This shows the peculiarity of $C_M^{\text{GIOs}}(\cdot)$ and the necessity of assumption  $C(\rho)=C(\rho^*)$.

%Our results in the note depends heavily on $C_M^{\text{GIOs}}(\cdot)$ which can be perceived as a convexity
%of the robustness of coherence.  In fact, $C_M^{\text{GIOs}}(\cdot)$ has a distinguished  property $$C_M^{\text{GIOs}}(\rho)\neq C_M^{\text{GIOs}}(\rho^*)\ \eqno{(20)}$$ for some $\rho$ and $M\in\Omega$, here $ \rho^*$ is the complex conjugate of $\rho$ However, for any coherence measure $C$, $$C(\rho)= C(\rho^*)\ \eqno{(21)}$$ is an axiomatic assumption proposed in \cite{Xu2} for studying
%coherence and imaginarity of quantum states which is a hot topic and recently receives much
%attention (see \cite{Xu2} and the references therein). On this basis, the axiomatic assumption $C(\rho)=C(\rho^*)$ is proposed. The (19)

%Now, let us turn to coherent state convertibility.
%

\vspace{0.1in} \section{Summary and discussion}

Among the most fundamental questions in  quantum coherence theory is state convertibility, it is aimed to study  whether
incoherent operations can introduce an order on the set of coherent states, i.e.,
whether, given two coherent states $\rho$  and $\sigma$, either $\rho$ can be transformed into $\sigma$ or vice versa.
Since the question of state convertibility in coherence resource theory is proposed \cite{Baumgratz},
understanding  exact
conditions for  existence of incoherent transformations
between coherent states has attracted a lot of work \cite{Streltsovade}.
In this work, we have determined exact conditions for coherence conversion under GIOs. Our conditions show that coherence measures from convexity  of the robustness of coherence are central. Based on these conditions, maximally incoherent states in a particular set are  classified.
This induces the majorization condition of determining the convertibility from pure states to mixed states under SIOs.
Furthermore, conditions of conversion between off-diagonal parts of coherent states
are also characterized. The study of state  convertibility for general resource theory has also been discussed recently \cite{Takagix, Datta}.

There still exist some interesting open questions. First, note that the existence proof of our Theorem 3.1 is not constructive,
given two states satisfying coherence order, a problem is how to construct desired GIOs realizing
convertibility? Second, can we offer an efficient algorithm to compute $C_{ M}^\text{GIOs}(\cdot)$?
Note that $C_{M}^\text{GIOs}(\cdot)$ is a generalization of quantum coherence fraction which quantifies the closeness between a given
state and the set of maximally coherent states \cite{Sun2}.
Therefore an efficient algorithm of $C_{ M}^\text{GIOs}(\cdot)$ is also efficient for
 quantum coherence fraction which is key in the framework of coherence theory.

%For any coherent state $\rho$, we write $\rho$ in the form $$\rho=\text{Re} \rho+i \text{Im}\rho,$$
%$\text{Re}(\rho)=\sum_{jk}(\text{Re}\rho_{jk})|j\rangle\langle k|$ is the real part of $\rho$, $\text{Im}(\rho)=\sum_{jk}(\text{Im}\rho_{jk})|j\rangle\langle k|$ is the imaginary part of $\rho$.
%For any coherence measure $C$, the non-negativity of $$C(\rho)-C(\text{Re}\rho)$$ quantifies the imaginary coherence \cite{Xu2}.
%Under axiomatic assumption $$C(\rho)=C(\rho^*),$$  it has been shown any measure $C$ in the framework of \cite{Baumgratz} has the property $C(\rho)-C(\text{Re}\rho)\geq 0$, here $ \rho^*$ is the complex conjugate of $\rho$ \cite{Xu2}.
%However, by the definition of robust cohernce, one can see $C_{\text {ROC}}(\rho)=C_{\text {ROC}}(\rho^*)$ which is also checked in \cite{Xu2}. By Theorem 1, we can obtain $\rho\xrightarrow{\text{GIO}} \rho^*$ and $\rho^*\xrightarrow{\text{GIO}} \rho$.
%Note that $$\text {GIOs} \subseteq \text{SIOs} \subseteq \text{IOs}\subseteq \text{MIOs}, \text{GIOs}\subseteq \text{DIOs},$$ hence $\rho\xrightarrow{\mathcal O} \rho^*$ and $\rho^*\xrightarrow{\mathcal O} \rho$ for $\mathcal O\in \{\text {SIOs,\ IOs,\ MIOs,\ DIOs}\}$.
%The monotonicity of coherence measure under $\mathcal{O}$ tells us $C(\rho)=C(\rho^*)$. Therefore the axiomatic assumption $C(\rho)=C(\rho^*)$ holds true actually.

%combining the properties (i) and (iv) of $C_M^{\text GIO}(\cdot)$,  one can see $C_M^{\text GIO}(\rho_p)\leq pC_M^{\text GIO}(|\psi^+\rangle\langle\psi^+|)$.

\vspace{0.1in}
{\it Acknowledgement---}
This research was supported by NSF of China
(12271452), NSF of Xiamen (3502Z202373018) and NSF of Fujian (2023J01028).

\vspace{0.1in}

{\it\bf Appendix: Proof of our results}\vspace{0.1in}
\vspace{0.1in}

Proofs of all results in this paper are given in the appendix.

Before giving the proof of our main results, we firstly recall some fundamental properties of GIOs. In fact, the notion of GIOs is equivalent to the
Schur channels \cite{Ckli,Paulsen,Water}. Suppose $\Phi$ is trace-preserving completely positive maps on density operators, the following statements are equivalent:

(1) $\Phi$ is a GIO, i.e., a Schur channel;

(2) $\Phi$ preserves incoherent basis states, i.e., $\Phi(|i\rangle\langle i|)=|i\rangle\langle i|$ for all $i$;

(3) For every Kraus representation of $\Phi(\rho)=\sum_{j}K_j \rho K_j^\dag,$ all Kraus operators $\{K_j\}$ are diagonal;

(4) $\Phi$ can be written as a Schur product form: $\Phi(\rho)=\tau\circ \rho$, where the matrix $\tau$ is positive semidefinite such that its diagonals are all equal to $1$, and the Schur product is denoted by $\tau\circ\rho=(\tau_{ij}\rho_{ij})$.

\vspace{0.1in}

{\bf Proof of properties of $C_M^\text{GIOs}(\rho)$.}

$ \begin{array}{ll}
(i)\ C_{ M}^\text{GIOs}(\rho)&=\max_{\Phi\in\text{ GIOs}} {\rm tr}(\Phi(\rho)M)-\frac{1}{d}\\
                        &=\max_{\tau\geq 0, \{\tau_{ii}=1\}_{i=1}^d} {\rm tr}((\rho_{ij}\tau_{ij})M)-\frac{1}{d}\\
                        &=\max_{\tau\geq 0, \{\tau_{ii}=1\}_{i=1}^d}\sum_{i\neq j}\rho_{ij}\tau_{ij}M_{ji}\end {array}.$

It is evident if $\rho\in{\mathcal I}$, then $$\rho_{ij}=0\  (1\leq i\neq j\leq d)$$ and so $C_{ M}^\text{GIOs}(\rho)=0$.
Otherwise, choosing $\tau_{ij} (i\neq j)$ such that $\rho_{ij}\tau_{ij}M_{ji}=|\rho_{ij}\tau_{ij}M_{ji}|$ and $\tau_{ii}=1$, then  $C_{ M}^\text{GIOs}(\rho)\geq 0$.

(ii) Note that the composition of two GIOs is also a GIO, monotonicity of $C_M^\text{GIOs}(\cdot)$ under all GIOs is evident.

(iii) Combining Theorem 1 \cite{Cktan} and $\text{GIOs}\subseteq\text{IOs}$, we get the desired.

(iv)  It is easy to check that $C_M^\text{GIOs}(\cdot)$ is non-increasing under mixing of quantum states.

(v) By (i), $$C_{ M}^\text{GIOs}(\rho)=\max_{\tau\geq 0, \{\tau_{ii}=1\}_{i=1}^d}\sum_{i\neq j}\rho_{ij}\tau_{ij}M_{ji}.$$ Combining  $\tau\geq 0$ with $\tau_{ii}=1 (1\leq i\leq d)$, we have $|\tau_{ij}|\leq 1$. Therefore
$$\sum_{i\neq j}\rho_{ij}\tau_{ij}M_{ji}\leq \sum_{i\neq j}|\rho_{ij}||M_{ji}|.$$ It is evident that $$C_{ M}^\text{GIOs}(\rho)\leq C_{l_1}(\rho)\max_{1\leq i\neq j\leq d}\{|M_{ij}|\}.$$ Choosing $\tau$ as the  Lemma 1 \cite{Piani}, a direct computation shows
$$\frac{C_{l_1}(\rho)}{d-1}\min_{1\leq i\neq j\leq d}\{|M_{ij}|\} \leq C_{ M}^\text{GIOs}(\rho).$$

(vi) By the form of $M$ and Theorem 2 \cite{Sun2},  (11) and (12) can be obtained.

\vspace{0.1in}

{\bf Proof of Theorem 3.1.} $``\Rightarrow"$ Assume  there exists a GIO $\Phi$ with $\Phi(\rho)=\sigma$, by the monotonicity of $C_M^\text{GIOs}(\rho)$ under GIOs,  we have $C_M^\text{GIOs}(\sigma)\leq C_M^\text{GIOs}(\rho)$.
Note that  every GIO is a Schur channel, i.e., $\Phi(\rho)=\tau\circ \rho$, thus $\rho_{ii}=\sigma_{ii}\ (i=1,2,\cdots, d)$.

 $``\Leftarrow"$  %The monotonicity and convexity are evident, we need only show that $C_{|\psi^+\rangle\langle \psi^+|}(\cdot)$ is non-negative and
%$C_{|\psi^+\rangle\langle \psi^+|}(\rho)=0$ for every $\rho\in {\mathcal I}$. If $\rho\in{\mathcal I}$, then $\Phi(\rho)\in{\mathcal I}$. It is easy to check
%$\rm tr(\Phi(\rho)|\psi^+\rangle\langle \psi^+|)=\frac{1}{d}$, so
%$C_{|\psi^+\rangle\langle \psi^+|}(\rho)=0$. For any coherent state $\rho=(\rho_{ij})$, we choose positive semidfinite matrix $\tau$ such that  its diagonals are all equal to $1$ and  $\rho_{ij}\tau_{ij}\geq 0\ (1\leq i,j\leq d$).
%By the correspondence of GIO and positive semidefinite matrix, we get  $\Phi_0(\rho)=\tau\circ \rho$. A direct computation shows $\rm tr((\Phi_0(\rho)|\psi^+\rangle\langle \psi^+|))\geq \frac{1}{d}$. Thus $C_{|\psi^+\rangle\langle \psi^+|}(\rho)\geq 0$. By the definition of $C_{|\psi^+\rangle\langle \psi^+|}(\cdot)$, it is easy to see that $C_{|\psi^+\rangle\langle \psi^+|}(\rho)=C_{U|\psi^+\rangle\langle \psi^+|U^\dag}(\rho)$ for any diagonal unitary matrix.
%Using Theorem 2 of \cite{Sun2}, we obtain $$C_{\text{ROC}}(\sigma)\leq C_{\text{ROC}}(\rho)\Leftrightarrow
By the definition of $C_{M}^\text{GIOs}(\rho)$, it is easy to see that $C_{M}^\text{GIOs}(\rho)=C_{U M U^\dag}^\text{GIOs}(\rho)$ for any diagonal unitary matrix. Therefore $$\max_{\Phi\in \text{GIOs}} \rm tr(\Phi(\rho)U M U^\dag)\geq  \rm tr(\sigma UMU^\dag).$$   This implies $$ \min_{M\in\Omega}\max_{\Phi\in \text{GIOs}} \rm tr((\Phi(\rho)-\sigma)UMU^\dag)\geq 0,$$ here the optimization is over all convex combintation of maximally coherent states. Note that GIOs is compact and convex, and $\Omega$ is convex, by the fundamental  von Neumann's minimax theorem \cite{Sion}, $$ \max_{\Phi\in \text{GIO}}\min_{M\in\Omega} \rm tr((\Phi(\rho)-\sigma)U M U^\dag)\geq 0.$$
Thus there exists a GIO $\Phi_0$ such that $$\rm tr((\Phi_0(\rho)-\sigma)UMU^\dag)\geq 0$$ for all $M\in\Omega$. In particular, we have
$$\rm tr((\Phi_0(\rho)-\sigma)U|\psi^+\rangle\langle \psi^+|U^\dag)\geq 0$$ for all diagonal unitary matrices $U$.
%From $C_M^{\text GIO}(\rho)\geq C_M^{\text GIO}(\sigma)$  for any  $M\in {\Omega}$, it is evident
%$$C_{|\psi^+\rangle\langle \psi^+|}(\rho)\geq C_{|\psi^+\rangle\langle \psi^+|}(\sigma).$$
%Therefore
In the following,we will show $\Phi_0(\rho)=\sigma$ by the mathematical induction.
For induction step, it is firstly shown that  $\Phi_0(\rho)=\sigma$ for a three-level system.
Secondly, we deduce a $l$ dimensional system satisfies the assertion by assuming a $l-1$ dimensional system does.
Let  $$\Phi_0(\rho)-\sigma=\left(\begin{array}{ccc}
                                0 & a_{12}+ib_{12} & a_{13}+ib_{13}\\
                                 a_{12}-ib_{12}  & 0 & a_{23}+ib_{23}\\

                              a_{13}-ib_{13}& a_{23}-ib_{23}& 0\end{array}\right),$$ $a_{ij}, b_{ij}$ are all real numbers, and $U=\left(\begin{array}{ccc}
                                e^{i\theta_1} & 0& 0\\
                                0  &  e^{i\theta_2}  & 0\\
                              0& 0&  e^{i\theta_3} \end{array}\right)$, a direct computation shows
$$\begin{array}{ll}
\text{Re}(a_{12}+ib_{12})e^{i(\theta_2-\theta_1)}+\text{Re}(a_{13}+ib_{13})e^{i(\theta_3-\theta_1)}+&\\
\text{Re}(a_{23}+ib_{23})e^{i(\theta_3-\theta_2)}\geq 0.&\end{array}$$
That is $$\begin{array}{ll}a_{12}\cos (\theta_2-\theta_1)-b_{12}\sin (\theta_2-\theta_1)+a_{13}\cos (\theta_3-\theta_1)-&\\b_{13}\sin (\theta_3-\theta_1)+
                                   a_{23}\cos (\theta_3-\theta_2)-b_{23}\sin (\theta_3-\theta_2)\geq 0.& \end{array}$$
Choosing $\theta_1=\theta_2=\theta_3=0$, we have $$a_{12}+a_{13}+a_{23}\geq 0. \ \ \ (\text A1)$$ Picking $(\theta_1, \theta_2, \theta_3)=(0, \pi, \pi)$, we can obtain $$ a_{23}-a_{12}-a_{13}\geq 0.  \ \ \ (\text A2)$$  Selecting $(\theta_1, \theta_2, \theta_3)=(0, 0, \pi)$, we get
$$a_{12}-a_{13}-a_{23}\geq 0. \ \ \ (\text A3)$$ Let $(\theta_1, \theta_2, \theta_3)=(0, \pi, 0 )$, we have
$$-a_{12}+a_{13}-a_{23}\geq 0. \ \ \ (\text A4)$$ It is evident that $$(\text A1)+(\text A2)\Rightarrow a_{23}\geq 0,$$
$$(\text A1)+(\text A3)\Rightarrow a_{12}\geq 0,$$ $$(\text A1)+(\text A4)\Rightarrow a_{13}\geq 0.$$
A direct computation shows that
$$(A2)+(A3)\Rightarrow a_{13}\leq 0,$$ $$(A2)+(A4)\Rightarrow a_{12}\leq 0,$$ $$(A3)+(A4)\Rightarrow a_{23}\leq 0.$$
Therefore $a_{12}=a_{13}=a_{23}=0$. Analogously,  we can also obtain
$$b_{12}=b_{13}=b_{23}=0,$$ and so $\Phi_0(\rho)=\sigma$.

Let $U=\sum_{i=1}^l e^{{\rm i} \theta_i}|i\rangle\langle i|$, $\Phi_0(\rho)-\sigma=(a_{ij}+{\rm i}b_{ij})$, here both $a_{ij}$ and $b_{ij} (1\leq i\neq j\leq l)$ are real numbers. we assume $$\rm tr(U^\dag(\Phi_0(\rho)-\sigma)U|\psi^+\rangle\langle \psi^+|)\geq 0. \ \ \ (\text A5)$$ A direct computation shows that condition (A5) is equivalent to  $$\sum_{1\leq i< j\leq l}(a_{ij}\cos(\theta_j-\theta_i)-b_{ij}\sin(\theta_j-\theta_i))\geq 0.\ \ \ (\text A6)$$ It is easy to see
$$\begin{array}{ll}
\sum_{1\leq i<j\leq l}(a_{ij}\cos(\theta_j-\theta_i)-b_{ij}\sin(\theta_j-\theta_i))&\\
=\sum_{1\leq i< j\leq l-1}(a_{ij}\cos(\theta_j-\theta_i)-b_{ij}\sin(\theta_j-\theta_i))+&\\
\sum_{1\leq i<l}(a_{il}\cos(\theta_l-\theta_i)-b_{il}\sin(\theta_l-\theta_i))\geq 0.&\end{array}$$
By the arbitrariness of $\theta_l$, substituting $\theta_l$ for $\pi+\theta_l$, we have
$$\begin{array}{ll}
\sum_{1\leq i< j\leq l-1}(a_{ij}\cos(\theta_j-\theta_i)-b_{ij}\sin(\theta_j-\theta_i))-&\\
\sum_{1\leq i<l}(a_{il}\cos(\theta_l-\theta_i)-b_{il}\sin(\theta_l-\theta_i))\geq 0.&\end{array}$$
Therefore $$\sum_{1\leq i< j\leq l-1}(a_{ij}\cos(\theta_j-\theta_i)-b_{ij}\sin(\theta_j-\theta_i))\geq 0.$$
By our induction, we have $a_{ij}=b_{ij}=0 (1\leq i\neq j\leq l-1)$. This implies
$$\sum_{1\leq i<l}(a_{il}\cos(\theta_l-\theta_i)-b_{il}\sin(\theta_l-\theta_i))=0.\ \ \ (\text A7)$$
Choosing $\theta_1=\theta_2=\ldots=\theta_{l}=0$ in (A7), we can obtain $$\sum_{1\leq i<l}a_{il}=0. \ \ \ (\text A8)$$
Picking $\theta_1=\pi, \theta_2=\theta_3=\ldots=\theta_{l-1}=0, \theta_l=\pi$ in (A7), one has $$a_{1l}-\sum_{2\leq i<l}a_{il}=0.\ \ \ (\text A9)$$
It is evident that $$(A8)+(A9)\Rightarrow a_{1l}=0.$$ Similarly, $a_{2l}=a_{3l}=\ldots=a_{l-1 l}=0$, and so $$\sum_{1\leq i<l} b_{il}\sin(\theta_l-\theta_i)=0 \ \ \ (\text A10)$$ from (A7).
Using  analogous treatments, we also have $b_{1l}=b_{2l}=\ldots=b_{l-1 l}=0$. The proof is completed.

\vspace{0.1in}

The proof of Theorem 3.2  depends on Theorem 3.3, so we firstly give the proof of Theorem 3.3.

\vspace{0.1in}

{\bf Proof of Theorem 3.3.} Assume $\rho=(\rho_{ij})$, by Theorem 3.1, we need only to prove $$C_M^\text{GIOs}(\rho)\leq C_M^\text{GIOs}(|\psi\rangle\langle\psi|),\ |\psi\rangle=\sum_{i=1}^d\sqrt{\rho_{ii}}|i\rangle.$$
From the proof of property (i) of $C_M^\text{GIOs}(\rho)$, we have
$$C_{ M}^\text{GIOs}(\rho)=\max_{\tau\geq 0, \{\tau_{ii}=1\}_{i=1}^d}\sum_{1\leq i\neq j\leq d}\tau_{ij}\rho_{ij}M_{ji}.$$
Similarily $$C_M^\text{GIOs}(|\psi\rangle\langle\psi|)=\max_{\tau\geq 0, \{\tau_{ii}=1\}_{i=1}^d}\sum_{1\leq i\neq j\leq d}\tau_{ij}\sqrt{\rho_{ii}\rho_{jj}}M_{ji}.$$
We divide the proof into two cases.

Case 1.  All $\rho_{ii}\neq 0 \ (i=1,2,\cdots, d)$.

Write $$C_{ M}^\text{GIOs}(\rho)=\max_{\tau\geq 0, \{\tau_{ii}=1\}_{i=1}^d}\sum_{1\leq i\neq j\leq d}\tau_{ij}\frac{\rho_{ij}}{\sqrt{\rho_{ii}\rho_{jj}}}\sqrt{\rho_{ii}\rho_{jj}}M_{ji},$$
then the $(i,j)$ position of ${\tau\circ\tau_0}$ is $\tau_{ij}\frac{\rho_{ij}}{\sqrt{\rho_{ii}\rho_{jj}}}$, here
$$\tau_0=\left(\begin{array}{cccc}
          1 & \frac{\rho_{12}}{\sqrt{\rho_{11}\rho_{22}}}& \cdots& \frac{\rho_{1d}}{\sqrt{\rho_{11}\rho_{dd}}}\\
          \frac{\rho_{21}}{\sqrt{\rho_{11}\rho_{22}}} & 1 & \cdots & \frac{\rho_{2d}}{\sqrt{\rho_{22}\rho_{dd}}}\\
          \vdots & \vdots &\ddots &\vdots \\
          \frac{\rho_{d1}}{\sqrt{\rho_{11}\rho_{dd}}}& \frac{\rho_{d2}}{\sqrt{\rho_{22}\rho_{dd}}}& \cdots&  1\end{array}\right),$$ and $\circ$ denotes the Schur product. Note that $$\tau_0=\left(\begin{array}{cccc}
          \frac{1}{\rho_{11}} & \frac{1}{\sqrt{\rho_{11}\rho_{22}}}& \cdots& \frac{1}{\sqrt{\rho_{11}\rho_{dd}}}\\
          \frac{1}{\sqrt{\rho_{11}\rho_{22}}} & \frac{1}{\rho_{22}} & \cdots & \frac{1}{\sqrt{\rho_{22}\rho_{dd}}}\\
          \vdots & \vdots &\ddots &\vdots \\
          \frac{1}{\sqrt{\rho_{11}\rho_{dd}}}& \frac{1}{\sqrt{\rho_{22}\rho_{dd}}}& \cdots&  \frac{1}{\rho_{dd}}\end{array}\right)\circ\rho\geq 0,$$
   this is due to the fact the Schur product of two positive semidefinite is also positive semidefinite \cite{Horn}. Hence $${\tau\circ\tau_0}\geq 0.$$ This implies $$C_M^\text{GIOs}(\rho)\leq C_M^\text{GIOs}(|\psi\rangle\langle\psi|).$$

Case 2.  $\rho_{ii}=0$ for some $i$.

For clarity, we firstly treat the qutrit case with $\rho_{22}=0$. It is easy to see $\rho_{12}=\rho_{23}=0$.
Choosing $$\tau_0=\left(\begin{array}{ccc}
            1 & 0 &\frac{\rho_{13}}{\sqrt{\rho_{11}\rho_{33}}}\\
            0 & 1 & 0\\
            \frac{\rho_{31}}{\sqrt{\rho_{11}\rho_{33}}}& 0 & 1\end{array}\right),$$ the $(1,3)$ and (3,1) positions of $\tau\circ\tau_0$ have desired property as the case 1. Hence $$C_M^\text{GIOs}(\rho)\leq C_M^\text{GIOs}(|\psi\rangle\langle\psi|).$$
            For the general case, we can choose $\tau_0$ as follows:

            (1) Non-diagonal elements of the $ith$ row and the $ith$ column are all $0$;

       (2) All diagonal elements are $1$;

       (3) Other entries are defined as the case 1.

       It is easy to check that such $\tau_0$ has the property as the case
        1. Therefore   $$C_M^\text{GIOs}(\rho)\leq C_M^\text{GIOs}(|\psi\rangle\langle\psi|).$$

\vspace{0.1in}

Based on Theorem 3.3, we can prove Theorem 3.2.

\vspace{0.1in}

 {\bf Proof of Theorem 3.2.}   Assume $$(|\psi_1|^2, \cdots, |\psi_d|^2 )^t\prec (\sigma_{11}, \cdots,\sigma_{dd})^t,$$ then there exists a SIO
 $\Phi_1$ such that $$\Phi_1(|\psi\rangle\langle\psi|)=|\eta\rangle\langle\eta|, |\eta\rangle=\sum_{i=1}^d\sqrt\sigma_{ii}|i\rangle$$ \cite{Chitambar2}. By Theorem 3.3, there exists some GIO
 $\Phi_2$ such that $\Phi_2(|\eta\rangle\langle\eta|)=\sigma$. Let $\Phi$ be the composition of $\Phi_1$ and $\Phi_2$, it is easy to see that $\Phi$ is a SIO and $\Phi(|\psi\rangle\langle\psi|)=\sigma$.

%{\it For $|\psi\rangle=\sum_{i=1}^d\psi_i|i\rangle$, $\sigma=(\sigma_{ij})$, $$(|\psi_1|^2, \cdots, |\psi_d|^2 )^t\prec (\sigma_{11}, \cdots,\sigma_{dd})^t \Rightarrow|\psi\rangle \xrightarrow{\text{SIO}}\rho\ \eqno{(15)}$$ here $\prec$ denotes the majorization relation between probability vectors.}
\vspace{0.1in}

{\bf Proof of Theorem 3.4.} By the compactness of DIO and MIO, the sufficiency can be followed from the proof of Theorem 3.1. For  the  necessity, we  claim that $$\Phi^{\dagger}(M)_{ii}=\frac{1}{d}$$  for $\Phi\in \text{MIO}$. Indeed, for arbitrary state $\tau$, we have
$$\rm{tr}(\Phi^{\dagger}(M)\Delta(\tau))=\rm{tr}(M\Phi(\Delta(\tau))=\frac{1}{d}.$$ Thus $\Phi^{\dagger}(M)_{ii}=\frac{1}{d}$. Now, assume that $$\Phi_0(\rho)-\triangle (\Phi_0(\rho))=\sigma-\triangle(\sigma) $$ for some $\Phi_0\in \mathcal O$. Then
$$\begin{array}{ll}
	& C_M^{\mathcal O}(\rho)\geq C_M^{\mathcal O}(\Phi_0(\rho)\\
	= &\max_{\Phi\in \mathcal{O}} \rm tr(\Phi(\Phi_0(\rho))M)-\frac{1}{d}\\
	=&
	\max_{\Phi\in \mathcal{O}} \rm tr(\Phi_0(\rho)\Phi^{\dagger}(M))-\frac{1}{d}\\
	=&\max_{\Phi\in \mathcal{O}} \rm tr(\sigma\Phi^{\dagger}(M))-\frac{1}{d}=C_M^{\mathcal O}(\sigma).\end{array}$$

In the following, we give an example to show $$C_M^{\text{GIOs}}(\rho)\neq C_M^{\text{GIOs}}(\rho^*).$$

\vspace{0.1in}

{\bf Example.}  Taking  $$M=\frac{1}{3}U_1|\psi^+\rangle\langle\psi^+|U_1^\dag+\frac{2}{3}U_2|\psi^+\rangle\langle\psi^+|U_2^\dag,$$ here $U_1=\text{diag}(1,\frac{3+4i}{5},1)$ and $U_2=\text{diag}(1, 1,\frac{3+4i}{5})$. That is
$M=(M_{ij})$ with $$\begin{array}{l}
M_{12}=\frac{13}{45}-\frac{4}{45}i,\\
M_{13}=\frac{11}{45}-\frac{8}{45}i,\\
M_{23}=\frac{1}{5}-\frac{4}{45}i,\\
M_{11}=M_{22}=M_{33}=1.\end{array}$$
One can check that
$$\begin{array}{ll}
&	C_M^{\text{GIOs}}(M) = \max_{\Phi\in \text{GIOs}} \rm tr(\Phi(M)M)-\frac{1}{3}\\
=& \max_{\tau\geq 0, \tau_{ii}=1}\sum_{i\neq j}\tau_{ij}|M_{ji}|^2=\sum_{i\neq j}|M_{ji}|^2\end{array}.$$
$$\begin{array}{ll}
	&	C_M^{\text{GIOs}}(M^*) = \max_{\Phi\in \text{GIOs}} \rm tr(\Phi(M^*)M)-\frac{1}{3}\\
	=& \max_{\tau\geq 0, \tau_{ii}=1}\sum_{i\neq j}\tau_{ij}(M_{ji})^2\\
	< &\sum_{i\neq j}|M_{ji}|^2.\end{array}$$
The last strict inequality holds true because $\max_{\tau\geq 0, \tau_{ii}=1}\sum_{i\neq j}\tau_{ij}(M_{ji})^2
\leq \sum_{i\neq j}|M_{ji}|^2$ and the equation holds true iff $(M_{ji}^2)=U(|M_{ji}|^2)U^{\dag}$ for some diagonal unitary $U$. To show this, one only need a direct computation saying $(M_{12}M_{23}\overline{M_{13}})^2$ is not a real number.

\end{document}